\documentclass[journal]{IEEEtran}

\usepackage{graphicx}
\usepackage{amsmath}
\usepackage{cite}
\usepackage{multirow}
\usepackage{amsfonts,amssymb}
\usepackage{flushend, cuted}
\usepackage{lipsum}
\usepackage{stfloats}
\usepackage{cases}
\usepackage{multirow}
\usepackage{booktabs}
\usepackage{balance}
\usepackage{enumitem}
\usepackage{inconsolata}
\usepackage{bbm}
\usepackage{color}
\usepackage{makecell}
\usepackage{url}
\usepackage[usenames,dvipsnames]{xcolor}
\usepackage{color}

\usepackage[ruled,linesnumbered]{algorithm2e}
\ifCLASSINFOpdf

\else

\fi
\begin{document}
\title{Communication-Free Distributed Charging Control for Electric Vehicle Group}
\author{Heyang~Yu,~\IEEEmembership{Graduate~Student~Member,~IEEE},
				Chuanzi~Xu,
				Weifeng~Wang,\\
		        Guangchao~Geng,~\IEEEmembership{Senior Member,~IEEE},
		        and~Quanyuan~Jiang,~\IEEEmembership{Senior Member,~IEEE}
\thanks{This work was supported by the project of the State Grid Zhejiang Electric Power Co. Ltd. under 5211HZ22001V. \emph{Corresponding author: Guangchao Geng}.
	
	 H. Yu, G. Geng and Q. Jiang are with the College of Electrical Engineering, Zhejiang University, Hangzhou, 310027, China (e-mail: $\{$yuheyang, ggc, jqy$\}$ @zju.edu.cn). C. Xu is with the Hangzhou Power Supply Company of State Grid Zhejiang Electric Power Co. Ltd., Hangzhou, 310016, China (e-mail: 389396733@qq.com). W. Wang is with the State Grid Zhejiang Electric Power Co. Ltd., Hangzhou, 310007, China (e-mail: 44126082@qq.com). 
	 }
}

\markboth{Accepted By IEEE Transactions on Smart Grid}%
{Shell \MakeLowercase{\textit{et al.}}: Bare Demo of IEEEtran.cls for IEEE Journals}
\maketitle
\begin{abstract}
The disordered charging of electric vehicles (EVs) in residential areas leads to a rapid increase of the peak load, causing transformer overload, but the charging control of EV group can effectively alleviate this phenomenon. However, existing charging control methods need reliable two-way communication infrastructure, which brings high operation costs and security risks. To offer a backup strategy for charging control of EVs after communication facilities fail, this paper proposes a communication-free charging control scheme to provide a decentralized on-site charging strategy for EV group. First, an uncontrollable EV group baseline estimation considering charging behaviors enabled by Gaussian mixture model (GMM) is proposed to acquire the capacity margin forecasting for controllable EVs. Next, this paper proposes a probabilistic distributed control method to assist users formulate the charging plan autonomously. Here, the charging behavior of EV group is regulated from an optimization with uncertain boundary conditions to a sampling with uncertain feasible regions expressed by a probability distribution. Finally, the scheme is verified via real-world EV charging data from a residential area in Hangzhou, China. The results show that this method can reduce the probability of transformer overload caused by out-of-order EV charging after a communication failure.
\end{abstract}

\IEEEpeerreviewmaketitle

\section*{Nomenclature} 
\subsection{Acronyms} 
\begin{IEEEdescription}[\IEEEusemathlabelsep\IEEEsetlabelwidth{$i, j, x, y$}]
	\item[AC] Alternating current.
	\item[ADMM] Alternating direction multiplier method.
	\item[AIC/BIC] Akaike/Bayesian information criterion.
	\item[CSMS] Charging station energy management system.
	\item[EM] Expectation maximization.
	\item[EV] Electric vehicle.
	\item[GMM] Gaussian mixture model.
	\item[LHS] Latin hypercube sampling.
	\item[MAC] Maximum acceptable capacity.
	\item[MCU] Microcontroller unit.
	\item[NLL] Negative log likelihood.
	\item[PVD] Peak-valley difference.
	\item[PC-IPM] Predictive Correction Interior Point Method.
	\item[PD-IPM] Primal Dual Interior Point Method.
	\item[QP] Quadratic programming.
	\item[TOU] Time-of-use. 
\end{IEEEdescription}
\vspace{-1.5em}
\subsection{Indices/Sets}
\begin{IEEEdescription}[\IEEEusemathlabelsep\IEEEsetlabelwidth{$i, j, x, y$}]
	\item[${\rm A}^{'}$] Sequence composed of $N_s$ randomly sampled $\alpha^{''}$ and $	y^{\alpha^{''}}$ that meet the conditions.
	\item[$h/\boldsymbol{H}$] Index/Set of  all equal time periods in one day, $\boldsymbol{H}: \{1, 2, ..., H\}$.
  	\item[$j$] Index of sampling EVs.
  	\item[$j_a$] Index of component in the optimal GMM.
  	\item[$k$] Index of Gaussian component.
	\item[$m/\boldsymbol{M}$] Index/Set of  EVs, $\boldsymbol{M}: \{1, 2, ..., M\}$.
	\item[$t/\boldsymbol{T}$] Index/Set of  time periods that EVs can be scheduled, $\boldsymbol{T}: \{1, 2, ..., T\}$.
	\item[$\boldsymbol{T}^{\rm{EV}}/\boldsymbol{T}^{\rm{EV}}_{j}$] Set of Baseline load of EV group/EV $j$.
  	\item[$\boldsymbol{T}^{\rm{load}}$] Set of day-ahead baseline load.
	\item[$\boldsymbol{T}^{\rm{res}}$] Set of daily forecasting of residential load.
	\item[$\boldsymbol{P}^{\rm{cha}}/\boldsymbol{P}^{\rm{st}}$] Probability set that EV is charing/of EV charging start time.
	\item[$\boldsymbol{P}^{\rm{st, nor}}$] Probability set after normalizing and intervalizing $\boldsymbol{P}^{\rm{st}}$.
  	\item[$\boldsymbol{Q}^{\rm{cha}}$] Set of charging power of controllable EVs in all schedule time periods.
  	\item[$\boldsymbol{y}$] Set of charging scheduling result of controllable EVs.
	\item[$\boldsymbol{\Gamma}_1$] Coefficient matrix of whether EV is charging.
  	\item[$\Psi$] Model of ternary GMM.
  	\item[$\rm Z$] Set of day-ahead capacity margin.
\end{IEEEdescription}
\vspace{-1.5em}
\subsection{Parameters}
\begin{IEEEdescription}[\IEEEusemathlabelsep\IEEEsetlabelwidth{$i, j, x, y$}]
	\item[$a/b$] Number of row/column of $\boldsymbol{\Gamma}_1$.
	\item[$C_1$] Maximum capacity of the transformer.
	\item[$H$] Number of all time periods in $\boldsymbol{H}$.
	\item[$K$] Number of components of GMM $\Psi$.
	\item[$M$] Number of EVs.
	\item[$N_s$] Number of elements in sequence ${\rm A}^{'}$.
	\item[$N^{\rm{total}}$] Number of residential users.
	\item[$Q^{\rm{r}}$] Fixed charing power.
	\item[$T$] Number of the maxium time periods that EVs can be scheduled in $\boldsymbol{H}$.
	\item[$T^{\rm{arr}}$] Arrival time of EV.
	\item[$X$] Number of highest consumption days within $Y$ similar days.
	\item[$Y$] Number of similar days.
\end{IEEEdescription}
\vspace{-1.5em}
\subsection{Variables}
\begin{IEEEdescription}[\IEEEusemathlabelsep\IEEEsetlabelwidth{$i, j, x, y$}]
	\item[$e_i$] Target charging energy (kWh) of record $i$.
	\item[$e_j$] Target charging energy (kWh) of EV $j$.
	\item[$K_{j_a}$] Optimal number of component  $j_a$ of GMM.
	\item[$p(\phi_i)$] Probability of a record of charging behavior $\phi_i$.
	\item[$Q^{\rm{cha}}_h$] Charging power of one controllable EV at time period $h$.
	\item[$Q^{\rm{cha}}_{m, h}$] Charging power of the $m$-th EV at time period $h$.
	\item[$Q_t^{\rm{load}}$] Total load of at time period $h$.
	\item[$Q_t^{\rm{res}}$] Residential load at time period $h$.
	\item[$T^{\rm{d}}/E^{\rm{d}}$] Departure time/energy demand after arrving.
	\item[$T^{\rm{EV}}_h/T^{\rm{EV}}_{j, h}$] Baseline load of EV group/EV $j$ at time period $h$.
  	\item[$T^{\rm{load}}_h$] Day-ahead baseline load at time period $h$.
	\item[$T^{\rm{res}}_h$] Daily forecasting of residential load at time period $h$.
	\item[$y$] Charging scheduling result of one controllable EV.
	\item[$P^{\rm{cha}}_h/P^{\rm{st}}_h$] Probability at time period $h$ that EV is charing/of EV charging start time.
	\item[$P^{\rm{st, nor}}_h$] Probability at time period $h$ in $\boldsymbol{P}^{\rm{st, nor}}$.
	\item[$y^{\alpha^{''}}$] Probability of randomly sampling.
	\item[$\alpha_i/\beta_i$] Charging start/duration time of record $i$.
	\item[$\alpha_j/\beta_j$] Charging start/duration time of EV $j$.
	\item[$\alpha^{'}/\beta^{'}$] Charging start/duration time of result.
	\item[$\alpha^{''}$] Time peroid of randomly sampling.
	\item[$\gamma_{a, b}$] Element of row $a$, column $b$ in $\Gamma_1$.
	\item[$\zeta_h$] Day-ahead capacity margin at time period $h$.
\end{IEEEdescription}
\vspace{-1.5em}
\subsection{Notations}
\begin{IEEEdescription}[\IEEEusemathlabelsep\IEEEsetlabelwidth{$i, j, x, y$}]
	\item[$\mathbb{R}$] Set of real numbers.
	\item[$N({\cdot})$] Gaussian distribution.
	\item[$U({\cdot})$] Uniform distribution.
	\item[$\mu^{k}/\Sigma_{k}$] Average and covariance matrix of Gaussian component $k$.
	\item[$\pi_{k}$] Mixing proportion of Gaussian component $k$.

\end{IEEEdescription}
\section{Introduction}
\IEEEPARstart{T}{he} promotion of widespread adoption of EVs as eco-friendly alternatives \cite{7786114} to conventional fuel-powered vehicles is a key step towards carbon neutrality. However, disordered charging of a large number of EVs in residential areas can lead to significant increases in peak load on the distribution system, causing transformer overload \cite{6919255}. This creates challenges for EV charging during peak periods, especially in residential areas wherever there are more multi-unit condominiums in China \cite{LI2023108969} and there are more single-family houses in the United States and Europe \cite{CROZIER2020114214}.

This negative impact of EVs on the power grid can be mitigated through the implementation of effective scheduling control strategies such as price-guided \cite{5484336} and optimization-based methods \cite{5356176}. Ref. \cite{2023022104, 2023022105} has shown that the TOU price-guided strategy or others can result in EVs charging simultaneously at the beginning of the low price period at night, causing occasional grid overloads. Optimization-based methods, on the other hand, need to solve a complex optimization problem. The central controller is generally deployed in the utility to solve centralized optimization problems. It needs to collect all kinds of information from the users and the power grid, and then send the control command to EVs after making a decision. Consequently, it requires a well-developed and reliable communication infrastructure to facilitate real-time information exchange and rapid execution of control commands.

However, the development of such an infrastructure can incur high deployment and operation costs and security risks. For instance, in the event of a communication failure, such as the loss of the link between the EV and the utility, the EV will be difficult to receive control commands and revert to disordered charging. Therefore, if we can provide a backup strategy for charging control of EVs after communication facilities fail, such as the centralized control method can be decentralized and distributed, then the probability of transformer overload caused by charging of EV out of control can be reduced.

Decentralized strategies have been proposed to decrease load peak. Ref. \cite{gan2012optimal, ardakanian2013distributed} propose decentralized algorithms to fill the load valley. Ref. \cite{9900304} proposes an EV CSMS based on blockchain, which aims to meet charging demands for large numbers of EVs. The existing decentralized strategies are converted from the original centralized algorithm such as using Benders decomposition \cite{7579628} and ADMM \cite{9112232}, which still need to iterate with the central controller in the process of decision-making. Therefore, if the communication infrastructure between the users and the utility fails at some time, these decentralized strategies are still difficult to take effect.

Many researchers have developed communication-free charging control strategies for EVs \cite{2015177, 6175816, 6672699, 6939355, 7275178, 8544481, 8903543, 9105414, 2023022101}. Ref. \cite{2015177} proposes a bidirectional frequency-dependent EV charge controller, which utilizes EV for offering frequency regulation service. In ref. \cite{6175816}, a rule-based algorithm is developed which allows PHEVs constructors to limit PHEVs charging current if they are charged at home. In \cite{6672699, 6939355}, some voltage-based charge controllers are discussed to smooth load profiles, reduce power losses and improve voltages. In \cite{7275178, 8544481, 8903543, 9105414, 2023022101}, S. Shafiq et al. have proposed more advanced control methods. Ref. \cite{7275178, 8544481, 8903543} have proposed autonomous EV charge control strategies to control EV charging in a way that mitigates their negative impacts. To ensure fair contribution from EVs connected to different nodes, the local nodal voltage and sensitivity to load changes are used as inputs to the proposed controller. In \cite{9105414, 2023022101},  researchers have used a machine learning-based algorithm to control EV charging load. These papers have done systematic work on solving voltage issues, which apply to EV charging in single-family houses equipped with distributed charging piles. In the scenario of above works, the voltage and sensitivity of different nodes (houses) will be significantly different. However, in the typical centralized parking lot of the residential community in China that contains EV group, the access points of different EVs are not far apart, so the voltage and sensitivity of each EV are similar and cannot be accurately calculated. Meanwhile, this work focuses more on the overload issues, where voltage may not be the problem, Therefore, the voltage- and sensivity-based method is not suitable for this scenario.

In light of the above discussion, facing the difficulty that the increased probability of transformer overload caused by the disordered charging of EV group in residential areas, and the difficulty of the existing centralized charging control methods requiring reliable and high operation cost communication facilities, this paper proposes a distributed charging control scheme as a backup to provide fully decentralized on-site charging strategy for EV group after a communication failure. First, an uncontrollable EV group baseline estimation considering charging behaviors modeled by GMM is proposed to acquire the capacity margin forecasting for controllable EVs. Next, a probabilistic distributed control algorithm for controllable EVs is proposed in which the charging behavior of the EV group is regulated from an optimization with uncertain boundary conditions to a sampling problem with the uncertain feasible regions expressed by a probability distribution so that the user can autonomously formulate the charging plan without the assistance of communication facilities. As a result,  in case of communication failure, the probability of the EV group charging out of control and the peak of the transformer in the residential area can be properly reduced.

It is highlighted with the following main contributions of this work:

\begin{enumerate}[label={(\alph*)}]
	\item A communication-free distributed charging control scheme containing hierarchical hardware architecture and algorithm design is proposed, in detail, it only needs a top-down broadcast capacity margin when the condition is normal and allows each EV to develop an on-site charging strategy autonomously as a backup after a communication failure.
	
	\item A probabilistic charging margin-based distributed control algorithm is designed, in which the optimization with uncertain boundary conditions is modeled as a sampling problem with the uncertain feasible regions expressed by the probability distribution. It reduces the probability of transformer overloading caused by out-of-order EV charging after a communication failure.
\end{enumerate}

The rest of the paper is organized as follow: Sec. \ref{2} states the problems, the basic idea and the charging control infrastructure for EV group in this work. Sec. \ref{3} introduces EV group baseline estimation method based on EV charging behavior. The communication-free distributed control algorithm is proposed in Sec. \ref{4}. Sec. \ref{5} offers case study and analysis via real-world data. Finally, Sec. \ref{6} concludes the paper.
\section{Problem and Framework}\label{2}
\subsection{Problem Statement}\label{2.1}
In this paper, the purpose of the load management of the residential EV group is to lower the load of the transformer in the residential area when the EV group is connected to a large number of charging piles for charging. Therefore, the charging control objective of the EV group is formulated to minimize the peak load of the transformer and minimize the PVD of load. As a consequence, the objective function can be expressed as (\ref{eq2023021501}) and (\ref{eq2023021502}) :

\begin{equation}
	\setlength{\abovedisplayskip}{3pt} 
	\setlength{\belowdisplayskip}{3pt}
		\left\{
		\begin{aligned}
		&\min\enskip\max\enskip \{Q_1^{\rm{load}}, Q_2^{\rm{load}}, ..., Q_t^{\rm{load}}\}\\
		&\max\enskip\min\enskip \{Q_1^{\rm{load}}, Q_2^{\rm{load}}, ..., Q_t^{\rm{load}}\}
		\end{aligned}
		\right.
		\label{eq2023021501}
\end{equation}
\begin{equation}
	\setlength{\abovedisplayskip}{3pt} 
	\setlength{\belowdisplayskip}{3pt}
	Q_t^{\rm{load}}=  \sum_{m = 1}^{M} Q^{\rm{cha}}_{m, t} + Q^{\rm{res}}_t, \quad t \in \boldsymbol{T} \subseteq \boldsymbol{H}\label{eq2023021502}
\end{equation}
where $\boldsymbol{T}$ is contained within the total period set $\boldsymbol{H}$.

\subsection{Basic Idea}
As we know, for the charging control of EV group in residential areas, on the one hand, a reliable communication channel from users to the utility (from bottom to top) is required to ensure that the utility can collect both load and user demand information, and make decisions on EV charging through optimization. On the other hand, the same reliable communication channel from the utility to the users (from top to bottom) is needed to ensure that the EV can obtain its corresponding charging instructions, so as to achieve ordered charging. Once one of the two communication channels fails, the utility will lose control of EV, which leads to the EV being in a state of disordered charging and an increased risk of transformer overload.

The following two basic ideas are proposed to address the issues:
\begin{enumerate}[label={(\alph*)}]
	\item First, a charging behavior-based group baseline is estimated and the capacity margin is calculated. Then the capacity margin is broadcast to users when the communication channel is available, which provides users a reference for subsequent charging decisions making.

	\item Second, when each EV is connected to charge, the local charging facility with an edge computing module calculates the probability that the EV can start charging in each period according to the demand information uploaded by EV and the capacity margin, samples the time that the EV can start charging and send the command to EV. Finally, each EV executes the command, which can reduce the probability of transformer overload.
\end{enumerate}
\subsection{Distributed Charging Infrastructure for EV Group}
\begin{figure*}[t] 
	\centering 
	\includegraphics[width=0.9\linewidth]{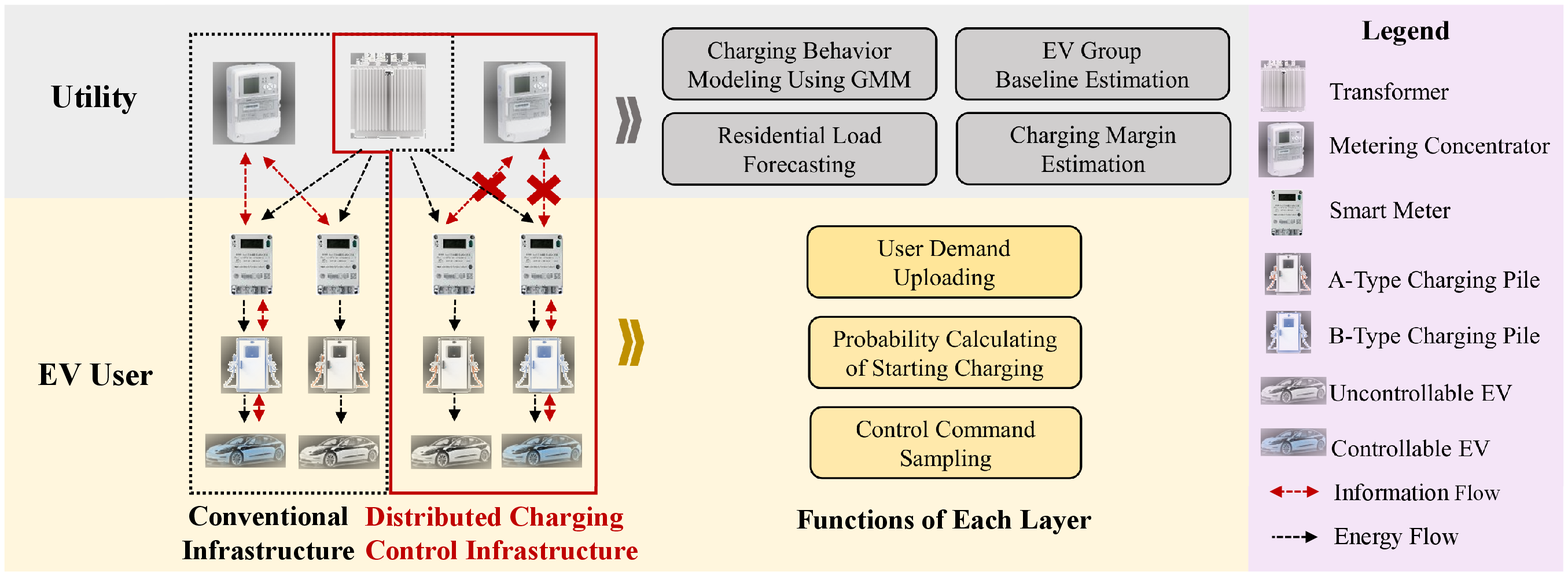} 
	\caption{Distributed charging control infrastructure for EV group} 
	\label{fig1} 
	\vspace{-1.0em}
\end{figure*}

In this paper, an overall distributed charging infrastructure for the EV group is proposed and shown in Fig. \ref{fig1}. The hardware architecture of this scheme consists of two levels: the utility and the EV user. They play different roles and complete their independent tasks through hierarchical cooperation.

\subsubsection{The Utility}
The utility includes the transformer and the metering concentrator. Generally, each distribution area under one transformer will have a metering concentrator, which is actually a connection point between the smart meter terminal and the grid. Its basic function is to read terminal data regularly, command transmission from the cloud system, data communication, etc. At this level, energy is collected to the transformer and data is collected to the metering concentrator. The metering concentrator is a central controller and is responsible for storing the historical charging behaviors of each EV, modeling the EV charging behavior, and then estimating the baseline load of the EV group. In addition, the metering concentrator is also responsible for forecasting the residential load and estimating the total baseline load by combining it with the baseline load of the EV group. After that, when the communication facility is available, the metering concentrator will collect the charging demand uploaded by each EV, and execute the optimization program in Sec. \ref{2.1} to make decisions on the EV charging period and power. If the communication facility fails, the capacity margin of this residential area can also be estimated and broadcast to each user from the metering concentrator. 
\subsubsection{The EV User}
The EV user includes a smart meter and a charging pile. When the communication facility is smooth, the EV user needs to execute the instructions of the concentrator through the charging pile. If not, the EV user needs to receive the estimated capacity margin of the residential area from the concentrator and make local decisions after getting the charging demand through a new edge computing module to determine the start time of charging after the EV arrives and implement the decisions through the charging pile. Generally speaking, this new edge computing module can be installed on both smart meters and charging piles. It is worth noting that the function of the traditional charging pile is to provide power for the EV. In this paper, we call it A-Type charging pile. The EV plugged in it is called an uncontrollable EV. In this paper, in addition to the original functions, the charging pile needs to be transformed to realize the interaction between the pile and the EV, and receive the demand information set by the user on the EV, such as the target charging energy and duration time, and transfer it to the edge computing module for charging decision. It is named B-Type charging pile. Meanwhile, the EV plugged in it is called a controllable EV.
\section{Charging Behavior Based Group Baseline Estimation}\label{3}
\subsection{Charging Behavior Modeling Enabled by GMM}
Uncontrollable or controllable EV loads are involved in EV charging loads. Therefore, modeling and estimating uncontrollable EV load accurately and reasonably is the key to effectively scheduling controllable EVs. One of the advantages of the Gaussian mixture model is that it can emulate any distribution and effectively represent the correlation between different variables using covariance, so it is used to model the joint distribution of charging behaviors of uncontrollable EVs \cite{3328313, 9509290}. Start time, duration time, and target charging energy which can be used to fully describe a complete record of charging behavior are selected in this work for modeling the joint distribution of charging behaviors \cite{3328313, 9509290}. 

Considering a historical charging dataset $\Phi$ with $N_1$ records of behavior, $\phi_i$ represents $i$-th record of charging behavior in $\Phi$, which can be expressed as (\ref{eq2023011701}):
\begin{equation}
	\setlength{\abovedisplayskip}{3pt} 
	\setlength{\belowdisplayskip}{3pt}
	\begin{aligned}
	{\Phi}: \{\phi_i=(\alpha_i, \beta_i, e_i): i \in N_1, \alpha_i \in \boldsymbol{H}, \beta_i \in \boldsymbol{H}, e_i \in \mathbb{R}\}\label{eq2023011701}
	\end{aligned}
\end{equation}

The probability $p(\phi_i)$ of a record of charging behavior $\phi_i$ can be approximated using ternary GMM $\Psi$ as (\ref{eq2022110201}):
\begin{equation}
\setlength{\abovedisplayskip}{3pt} 
\setlength{\belowdisplayskip}{3pt}
\Psi: p(\phi_i) = \sum_{k=1}^{K} {\pi_{k}N(\phi_i|\mu^{k},\Sigma_{k})} \label{eq2022110201}
\end{equation}

In this work, the parameters of (\ref{eq2022110201}) can be estimated using EM method. Moreover, AIC, BIC, or NLL can be used to determine the optimal $K_{j_a}$ number of components of GMM in $j_a$-th dimension. The optimal number $K$ can be calculated using (\ref{eq2023011702}):
\begin{equation}
	\setlength{\abovedisplayskip}{3pt} 
	\setlength{\belowdisplayskip}{3pt}
	K = \prod _{j_a = 1}^3 K_{j_a} \label{eq2023011702}
\end{equation}
\subsection{EV Group Baseline Estimation}
Baseline load refers to the electricity that users should consume without demand response or load curtailment \cite{8957454}. EV group baseline refers to the power that uncontrollable EVs. Therefore, the day-ahead baseline load set $\boldsymbol{T}^{\rm{load}}$ can be formulated as (\ref{eq2023011703}):
\begin{align}
	\setlength{\abovedisplayskip}{3pt} 
	\setlength{\belowdisplayskip}{3pt}
	\boldsymbol{T}^{\rm{load}} = \boldsymbol{T}^{\rm{res}} + \boldsymbol{T}^{\rm{EV}}\label{eq2023011703}
\end{align}

$\boldsymbol{T}^{\rm{load}}$, $\boldsymbol{T}^{\rm{res}}$ and $\boldsymbol{T}^{\rm{res}}$ are defined by (\ref{eq2023011704})-(\ref{eq2023011706}):
\begin{align}
	\setlength{\abovedisplayskip}{3pt} 
	\setlength{\belowdisplayskip}{3pt}
	\boldsymbol{T}^{\rm{load}}&: \{T^{\rm{load}}_h: h \in \boldsymbol{H}\} \label{eq2023011704}\\
	\boldsymbol{T}^{\rm{res}}&: \{T^{\rm{res}}_h: h \in \boldsymbol{H}\} \label{eq2023011705}\\
	\boldsymbol{T}^{\rm{EV}}&: \{T^{\rm{EV}}_h: h \in \boldsymbol{H}\} \label{eq2023011706}
\end{align}

Generally speaking, the calculation method of baseline load should be convenient and feasible enough to accept, calculate and implement. To be simple enough and obtain a conservative charging margin, $HighXofY$ which calculates the average load of the $X$ highest consumption days within $Y$ similar days is used to calculate the EV group baseline \cite{6775334}.  Therefore, after obtaining the GMM $\Psi$, we can generate the EV charging behavior of the next day by generating $N_1\over{X}$ records of data in GMM $\Psi$ for estimating baseline, where LHS is used for obtaining samples efficiently and comprehensively. Assuming that the dataset $\Phi_1$ is composed of $N_1\over{X}$ records of data and $\boldsymbol{T}^{\rm{EV}}_{j}: \{T^{\rm{EV}}_{j, h}: h \in \boldsymbol{H}\}$ is the charging power of $j$-th EV. $\boldsymbol{T}^{\rm{EV}}_{j}$ can be calculated by (\ref{eq2023011709}):
\begin{equation}
	\setlength{\abovedisplayskip}{3pt} 
	\setlength{\belowdisplayskip}{3pt}
	\left\{
	\begin{aligned}
		&T^{\rm{EV}}_{j, h} = 0, \forall h\in\left [1, \alpha_j - 1 \right)\cup\left (\alpha_j +\beta_j, H \right] \\
		&T^{\rm{EV}}_{j, h} = {e_j\over {\beta_j - \alpha_j + 1}}, \forall h\in\left [\alpha_j, \alpha_j +\beta_j - 1 \right] 
	\end{aligned}
	\right.
	\label{eq2023011709}
\end{equation}

Therefore, the EV group baseline can be calculated by (\ref{eq2023011710}):
\begin{equation}
	\setlength{\abovedisplayskip}{3pt} 
	\setlength{\belowdisplayskip}{3pt}
	T^{\rm{EV}}_{h}  =\sum_{j = 1}^{N_1\over{X}} T^{\rm{EV}}_{j, h}\label{eq2023011710}
\end{equation}

Finally, the overall EV group baseline estimation algorithm is shown in Algorithm \ref{alg:BLE}.
\begin{algorithm}[t]\small
	\caption{Baseline Load Estimation($\Phi$, $T^{\rm{res}}$)}
	\label{alg:BLE}
	\KwIn{A charging dataset $\Phi$ with $N_1$ records in $X$ days and a residential load forecasting daily $\boldsymbol{T}^{\rm{res}}$}
	\KwOut{A day-ahead load baseline $\boldsymbol{T}^{\rm{load}}$}
	Determine optimal number $K$ of GMM components\\
	Fit GMM joint distribution $\Psi$ of charging behavior using dataset $\Phi$\\
	Generating and sampling $N_1\over{X}$ charging records subject to GMM $\Psi$ using LHS\\
	Estimate day-ahead EV group baseline $\boldsymbol{T}^{\rm{EV}}$ using (\ref{eq2023011709}) and (\ref{eq2023011710})\\
	Forecast day-ahead load baseline $\boldsymbol{T}^{\rm{load}}$ using (\ref{eq2023011706})\\
\end{algorithm}
\section{Communication-Free Distributed Charging Control}\label{4}
In this section, when the communication infrastructure between the utility and the users fails, the global load will be unknown, the distributed charging control problem of the EV group is modeled as a sampling problem in the uncertain feasible decision space expressed by a probability distribution. 
\subsection{Control Objective and Feasible Region}
\subsubsection{Control Objective}
In this problem, the charging control objective of the EV group is to minimize the peak and PVD of the transformer load, that is, the EVs can be charged when the transformer load is at a low level, which is also called peak shaving and valley filling. 
\subsubsection{Feasible Region}
Meanwhile, in this problem, the feasible region of EV charging control is composed of charging start time, duration time, and charging power, where $\boldsymbol{y}$ is used to represent the result of charging scheduling, which can be expressed as (\ref{eq2023021001}):
\begin{equation}
	\setlength{\abovedisplayskip}{3pt} 
	\setlength{\belowdisplayskip}{3pt}
		\left\{
	\begin{aligned}
		&\boldsymbol{y}: \{y=(\alpha^{'}, \beta^{'}, \boldsymbol{Q}^{\rm{cha}}):  \alpha^{'} \in \boldsymbol{H}, \beta^{'}\in \boldsymbol{H}\}\\
		&\boldsymbol{Q}^{\rm{cha}} : \{Q^{\rm{cha}}_h: h \in\left [\alpha^{'}, \alpha^{'} +\beta^{'} - 1 \right] \}
	\end{aligned}
	\right.
	\label{eq2023021001}
\end{equation}
\subsection{Constraints}
In this section, we will introduce the detailed constraints that the method in this work needs to meet, including user demand constraints and constraints in the communication-free distributed scenario.
\subsubsection{User Demand Constraints}
Every time the users end using the EV and arrive at the parking lot, they need to input the demand information for the next use of the EV, including the energy demand $E^{\rm{d}}$ and the departure time $T^{\rm{d}}$. Therefore, $\boldsymbol{y}$ need to satisfy (\ref{eq2023021002})-(\ref{eq2023021004}):
\begin{align}
	\setlength{\abovedisplayskip}{3pt} 
	\setlength{\belowdisplayskip}{3pt}
	\alpha^{'} &\geq T^{\rm{arr}} \label{eq2023021002} \\
	\beta^{'} + \alpha^{'} - 1 &\leq T^{\rm{d}}\label{eq2023021003} \\
	\sum_{j = \alpha^{'}}^{\beta^{'} + \alpha^{'} - 1}Q^{\rm{cha}} &= E^{\rm{d}}\label{eq2023021004} 
\end{align}
\subsubsection{Communication-Free Scenario Constraints}
To ensure the high charging effciency of EVs and reduce the complexity of decision-making in the communication-free distributed charging scenario, $Q^{\rm{cha}}$ also need to satisfy (\ref{eq2023021005}) and (\ref{eq2023021006}):
\begin{align}
		\setlength{\abovedisplayskip}{3pt} 
		\setlength{\belowdisplayskip}{3pt}
	&Q^{\rm{cha}} = 0 , \forall h\in\left [T^{\rm{arr}}, \alpha^{'} - 1 \right)\cup\left (\alpha^{'} +\beta^{'} , T^{\rm{d}} \right] 	\label{eq2023021005}\\
	&Q^{\rm{cha}} = Q^{\rm{r}}, \forall h\in\left [\alpha^{'}, \alpha^{'} +\beta^{'}- 1 \right] \label{eq2023021006}
\end{align}
where $Q^{\rm{r}}$ is generally the rated power of the charging pile. Equation (\ref{eq2023021005}) shows the continuity of EV charging power in this scenario and reduces the complexity of the feasible region. Equation (\ref{eq2023021006}) ensures the high efficiency of the charging process of EV, which is different from the traditional method of charging with reducing power in this scenario. According to (\ref{eq2023021004}) and (\ref{eq2023021006}), the charging duration time of the EV can be calculated as (\ref{eq2023021007}):
\begin{equation}
	\setlength{\abovedisplayskip}{3pt} 
	\setlength{\belowdisplayskip}{3pt}
	\beta^{'} = {E^{\rm{d}}\over{Q^{\rm{r}}}}\label{eq2023021007} 
\end{equation}

As a result, according to the constraints of the communication-free scenario, the feasible region of EV charging control can be simplified, and only the charging start time is included.
\subsection{Uncertain Feasible Region}
In this problem, the load feasible region of charging control of EV group can be expressed by capacity margin, which refers to the EV load that can be connected to charging under the transformer without exceeding the capacity of the transformer. Therefore, the day-ahead capacity margin $\rm Z$ can be estimated by $\boldsymbol{T}^{\text{load}}$, defined and formulated as (\ref{eq2023011707}): 
\begin{equation}
	\setlength{\abovedisplayskip}{3pt} 
	\setlength{\belowdisplayskip}{3pt}
	\left\{
	\begin{aligned}
		&{\rm Z}: \{\zeta_h: h \in \boldsymbol{H}\}\\
		&{\rm Z} = C_1 - \boldsymbol{T}^{\rm{load}} 
	\end{aligned}
	\right.
	\label{eq2023011707}
\end{equation}
The uncertainty of the feasible region in this paper comes from the uncertainty of day-ahead load and capacity margin forecasting on the one hand, and the uncertainty of the arrival time of EVs on the other hand. In the communication-free scenario of distributed control, once the day-ahead load and capacity margin forecasting results are determined at the beginning of the day when the EV arrives, the feasible region (i.e. the optional charging start time) can be determined. The algorithm needs to choose the best charging start time. However, in this scenario, it is difficult for a single electric vehicle to obtain global demand information and transformer load, so probability distribution sampling can be used instead of online optimization to decide control command. The probability that EV is charging can be denoted and estimated by capacity margin $\rm Z$, which can be expressed as (\ref{eq2023021008}):
\begin{equation}
	\setlength{\abovedisplayskip}{3pt} 
	\setlength{\belowdisplayskip}{3pt}
	\left\{
	\begin{aligned}
		\boldsymbol{P}^{\rm{cha}} &= [P^{\rm{cha}}_h: h \in H]\\
		P^{\rm{cha}}_h &: {\zeta_h \over \sum_{h=1}^H \zeta_h}  
	\end{aligned}
	\right.
	\label{eq2023021008}
\end{equation}
 
According to the probability that EV is charging $\boldsymbol{P}^{\text{cha}}$, the probability of starting charging in each period  can be denoted by (\ref{eq2023021010}):
\begin{equation}
	\setlength{\abovedisplayskip}{3pt} 
	\setlength{\belowdisplayskip}{3pt}
	\boldsymbol{P}^{\rm{st}} = [P^{\rm{st}}_h: P^{\rm{st}}_h\geq0, h \in H]\label{eq2023021010}
\end{equation}

In accordance with the relationship between $\boldsymbol{P}^{\rm{st}}$ and $\boldsymbol{P}^{\rm{cha}}$, they satisfy (\ref{eq2022110204}):
\begin{algorithm}[t]\small
	\caption{Distributed Charging Control ($\boldsymbol{T}^{\rm{load}}$, $E^{\rm{d}}$, $T^{\rm{d}}$)}
	\label{alg:DCCM}
	\KwIn{A charging dataset $\Phi$ with $N$ records in $X$ days and a residential load forecasting daily $\boldsymbol{T}^{\rm{res}}$}
	\KwOut{A day-ahead load baseline $\boldsymbol{T}^{\rm{load}}$}
	For concentrator, calculate day-ahead capacity margin $\rm Z$ and the probability $\boldsymbol{P}^{\rm{cha}}$ that EV is charging through (\ref{eq2023011707}) and (\ref{eq2023021008}), and broadcast them to each smart meter once a day\\
	For smart meter connected the B-Type charging pile, when the EV arrive, collect the EV charging demand $E^{\rm{d}}$ and $T^{\rm{d}}$\\
	Calculate the duration time $\beta^{'}$ through (\ref{eq2023021007})\\
	Calculate the probability $\boldsymbol{P}^{\rm{st}}$ of EV charging start time through (\ref{eq2023021010}), (\ref{eq2023021011}) \\
	For variable $\alpha^{'}$,  randomly sample to obtain $N_s$ records and obtain a sequence ${\rm A}^{'}$ subject to distribution $\boldsymbol{P}^{\rm{st, nor}}$\\
	Randomly sample from the sequence ${\rm A}^{'}$ and acquire the charging start time $\alpha^{'}$\\
	Download the command $y = (\alpha^{'}, \beta^{'}, Q^{\rm{r}})$and start charing at $\alpha^{'}$
\end{algorithm}
\begin{equation}
	\setlength{\abovedisplayskip}{3pt} 
	\setlength{\belowdisplayskip}{3pt}
	\boldsymbol{\Gamma}_1\boldsymbol{P}^{\rm{st}} = \boldsymbol{P}^{\rm{cha}} \label{eq2022110204}
\end{equation}
where $\boldsymbol{\Gamma}_1=[\gamma_{a,b}: a, b \in H]$ is the coefficient matrix of whether the EV is charging, which form is shown in (\ref{eq2022110205})-(\ref{eq2022110206}):
\begin{equation}
	\setlength{\abovedisplayskip}{3pt} 
	\setlength{\belowdisplayskip}{3pt}
	\left\{
	\begin{aligned}
		\gamma_{a, b} =1,\enskip &1\leq b\leq a\\
		\gamma_{a, b} =1,\enskip &H-\beta^{'}+a+1\leq b\leq H\\
		\gamma_{a, b} = 0,\enskip &a<b<H-\beta^{'}+a+1\\
	\end{aligned}
	\right.
	\text{when} \enskip 1\leq a\leq\beta^{'}
	\label{eq2022110205}
\end{equation}
\begin{equation}
	\setlength{\abovedisplayskip}{3pt} 
	\setlength{\belowdisplayskip}{3pt}
	\left\{
	\begin{aligned}
		\gamma_{a, b} =1,\enskip &a-\beta^{'} +1\leq b\leq a\\
		\gamma_{a, b} = 0,\enskip &1\leq b<a-\beta^{'} +1\\
		\gamma_{a, b} = 0,\enskip &a<b\leq H\\
	\end{aligned}
	\right.
	\text{when} \enskip \beta^{'}< a\leq H
	\label{eq2022110206}
\end{equation}

The determinant form of $\boldsymbol{\Gamma}_1$ is shown in Appendix A. Equation (\ref{eq2022110204})-(\ref{eq2022110206}) can be solved by the least square method with constraints. The objective function of this problem can be expressed as (\ref{eq2023021011}): 
\begin{equation}
		\setlength{\abovedisplayskip}{3pt} 
		\setlength{\belowdisplayskip}{3pt}
		\min\quad {\|\boldsymbol{P}^{\rm{cha}} - \boldsymbol{\Gamma}_1\boldsymbol{P}^{\rm{st}}\|}^2 \label{eq2023021011}
\end{equation}
\begin{figure}[t] 
	\centering 
	\includegraphics[width=0.8\linewidth]{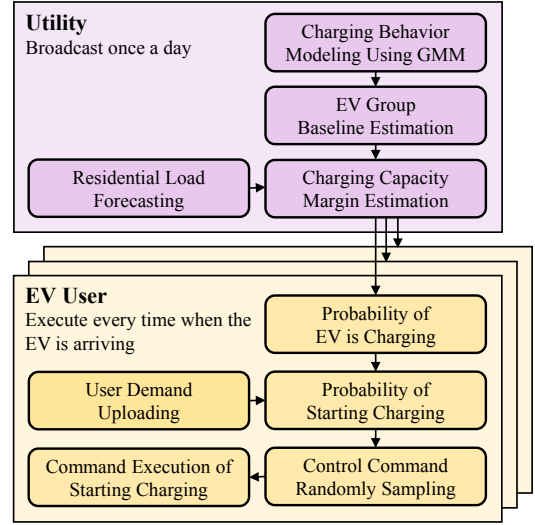} 
	\caption{Flowchart of the proposed method} 
	\label{fig10} 
	\vspace{-1.0em}
\end{figure}
The constraint of this problem is (\ref{eq2023021010}). It is able to prove that it is a QP problem with constraints and the problem is solvable (proved in Appendix B), and the problem is guaranteed to have an optimal solution. Because the problem needs to be solved in the local charging facility with an edge computing module when the computing resources of the module are sufficient, the problem can be solved using solvers; If the edge computing module is limited by the economic cost and the computing resources are insufficient, this problem can be solved using the PD-IPM \cite{vanderbei2020linear}.
\subsection{Commands Making Method}
For each EV, after obtaining the probability distribution of starting charging $\boldsymbol{P}^{\rm{st}}$ in each period, the starting time (i.e. control command) that conforms to the distribution can be sampled randomly, which includes the following steps:
\begin{enumerate}[label={(\alph*)}]
	\item According to (\ref{eq2023021002}) and (\ref{eq2023021003}), normalize and intervalize the probability distribution $\boldsymbol{P}^{\rm{st}}$ to obtain distribution $\boldsymbol{P}^{\rm{st, nor}}$, where it can be denoted and calculated as (\ref{eq2023021701}):
	\begin{equation}
		\setlength{\abovedisplayskip}{3pt} 
		\setlength{\belowdisplayskip}{3pt}
		\left\{
		\begin{aligned}
		&\boldsymbol{P}^{\rm{st, nor}} = [P^{\rm{st, nor}}_h: P^{\rm{st}}_h, h \in [T^{\rm{arr}},T^{\rm{d}}-\beta^{'} + 1]]\\
		&P^{\rm{st, nor}}_h =  {P^{\rm{st}}_h \over \sum_{h = T^{\rm{arr}}}^{T^{\rm{d}}-\beta^{'} + 1} {P^{\rm{st}}_h}} 
		\end{aligned}
		\right.
		\label{eq2023021701}
	\end{equation}	
	
	\item Acceptance-rejection sampling method \cite{flury1990acceptance} is used to acquire the sequence ${\rm A}^{'}$ that follows the distribution $\boldsymbol{P}^{\rm{st, nor}}$. For variable group $(\alpha^{'}_i, y^{\alpha^{'}}_i)$ subject to (\ref{eq2023021702}),  randomly sample to obtain $N_s$ records if $P^{\rm{st, nor}}_{\alpha^{'}_i} \geq y^{\alpha^{'}}_i$, and obtain a sequence ${\rm A}^{'}$ with $N_s$ data.
	\begin{equation}
		\setlength{\abovedisplayskip}{3pt} 
		\setlength{\belowdisplayskip}{3pt}
		\left\{                                                
		\begin{aligned}                                        
			&\alpha^{'}_i  \in \alpha^{''} \sim  U{(T^{\rm{arr}}, T^{\rm{d}}-\beta^{'} + 1)}\\
			&y^{\alpha^{'}}_i \in y^{\alpha^{''}} \sim  U{(0,\max(\boldsymbol{P}^{\rm{st, nor}}))}  
		\end{aligned}
		\right.
 		\label{eq2023021702}
	\end{equation}

	\item  Randomly sample from the sequence ${\rm A}^{'}$ and acquire the charging start time $\alpha^{'}$.
	
	\item  Download the command $y = (\alpha^{'}, \beta^{'}, Q^{\rm{r}})$ and start charing at $\alpha^{'}$.
\end{enumerate}

Finally, charging can be started. As a summary of the proposed communication-free distributed charging control for the EV group, the algorithm is provided as shown in Algorithm \ref{alg:DCCM}, and the flowchart of the proposed method is shown in Fig. \ref{fig10}.
\begin{table}[t]
	\renewcommand{\arraystretch}{1.2}
	\setlength\tabcolsep{2pt}
	\caption{Basic Parameters of Residential Area}
	\vspace{-1.5em}	
	\begin{center}
		\begin{tabular}{c c c c}
			\hline
			\hline
			Type &Parameters &Type &Parameters   \\
			\hline
			User Number & 100 & Rated Power of AC Pile & 7kW\cite{EVstandard1}   \\
			EV Penetration Rate & 60\% &Transformer Capacity & 600kVA\\
			Time Resolution & 15min& & \\
			\hline
			\hline
		\end{tabular}
		\label{table1}
	\end{center}
	\vspace{-1.5em}
\end{table}
\begin{table}[t]
	\renewcommand{\arraystretch}{1.2}
	\setlength\tabcolsep{12pt}
	\caption{Electricity Price of Residents in Hangzhou, China  \cite{2022022103}}
	\vspace{-1.5em}
	\begin{center}
		\begin{tabular}{c c c}
			\hline
			\hline
			Type &Time (h) &Price (CNY)\\
			\hline
			Peak Period&8:00-22:00  &0.568\\
			Valley Period&22:00-8:00 (Next Day)  &0.288\\
			\hline
			\hline
		\end{tabular}
		\label{table7}
	\end{center}
	\vspace{-1.5em}
\end{table}
\section{Case Study and Analysis}\label{5}
In this section, a series of case studies are conducted based on the real-world data of residential areas in Hangzhou, China, to verify the feasibility and advantages of this method.
\subsection{Simulation Setting}
\subsubsection{Parameter Setting and Scenario Genaration}
The basic parameters of a typical residential area located in Hangzhou, China are presented in Table \ref{table1}, where EV penetration rate $PR$ is denoted as (\ref{eq2023030901}): 
\begin{equation}
	\setlength{\abovedisplayskip}{3pt} 
	\setlength{\belowdisplayskip}{3pt}
	PR = {M \over N^{\rm{total}}}\label{eq2023030901}
\end{equation}
Here it is assumed that there is at most one EV in each family. Besides, to test the effectiveness of the proposed algorithm, there are three components of the scenarios: the type of communication link failure, the type of EV energy demand, and the level of residential load. The specific details on each of these components are provided in Table \ref{table5}. Two different demand types can show different test effects. By homogeneous, we mean that EVs have the same demand or action (all EVs plug in for charging at the same time, have the same deadline, and need to charge the same energy. By non-homogeneous, we mean that the demand or action is not necessarily the same for all EVs (EVs may plug in for charging at different times, have different deadlines, and charge different energy). Moreover, by comparing the performance of the proposed algorithm and the conventional orderly charging method across different residential load levels, the upper limit of different algorithms can be obtained.
\subsubsection{Simulation Environment and Tools}
In this paper, all the numerical simulation for algorithm validation is mainly carried out on a computer equipped with Intel Xeon X5650 2.67 GHz CPU and 24 GB RAM. All the charging optimization models can be solved by Ipopt 3.13.3. Python is used as a simulation platform for algorithm implementation.
\begin{table}[t]
	\renewcommand{\arraystretch}{1.2}
	\setlength\tabcolsep{2pt}
	\caption{Three Components of Test Scenarios}
	\vspace{-1.5em}	
	\begin{center}
		\begin{tabular}{c l}
			\hline
			\hline
			Component &Contents \\
			\hline
			\multirow{2}{*}{\makecell[c]{Communication \\Infrastructure Failure}} & Complete disconnection of  B-Type charging piles\\
			& Partial disconnection of B-Type charging piles\\
			\hline
			\multirow{2}{*}{\makecell[c]{EV Charging Demand}} & Homogeneous demand\\
			& Non-homogeneous demand\\
			\hline
			\multirow{3}{*}{\makecell[c]{Residential Load Level}} & Low level\\
			& Medium level\\
			& High level\\
			\hline
			\hline
		\end{tabular}
		\label{table5}
	\end{center}
\vspace{-1.5em}
\end{table}
\begin{table}[t]
	\renewcommand{\arraystretch}{1.2}
	\setlength\tabcolsep{7pt}
	\caption{EV Charging Demand Parameters}
	\vspace{-1.5em}
	\begin{center}
		\begin{tabular}{c c c}
			\hline
			\hline
			EV Demand Type &Parameters Type &Parameters   \\
			\hline
			\multirow{3}{*}{\makecell[c]{Homogeneous \\Demand}} &Arrival Time & $T^{\rm{arr}} =77$(15min) \\
			&Departure Time &  $T^{\rm{d}}=30$(15min)  \\
			&Charging Energy & $E^{\rm{d}} = 35$(kW·h)\\
			\hline
			\multirow{4}{*}{\makecell[c]{Non-Homogeneous \\Demand}} &Arrival Time & $T^{\rm{arr}}  \sim N(77, 8)$(15min)  \\
			&Departure Time & $T^{\rm{d}} \sim N(30, 4)$(15min)  \\
			&\multirow{2}{*}{\makecell[c]{Charging Energy}}  &\multirow{2}{*}{\makecell[c]{ Follow the distribution of \\Fig. 2(c) }} \\
			&&\\
			\hline
			\hline
		\end{tabular}
		\label{table4}
	\end{center}
\vspace{-1.5em}
\end{table}
\subsection{Result for Group Baseline Estimation}
In this section, the availability of the EV group baseline estimation method proposed in Sec. \ref{2} is verified. The charging data of EVs we use is the real data of a residential area in Hangzhou, China, in April 2021. Therefore, it is a natural non-homogeneous demand scenario. The charging piles in the residential area are all A-Type charging piles. Therefore, the charging start time, duration time, and charging power are modeled for testing.
\begin{figure*}[t] 
	\centering 
	\includegraphics[width=1.0\linewidth]{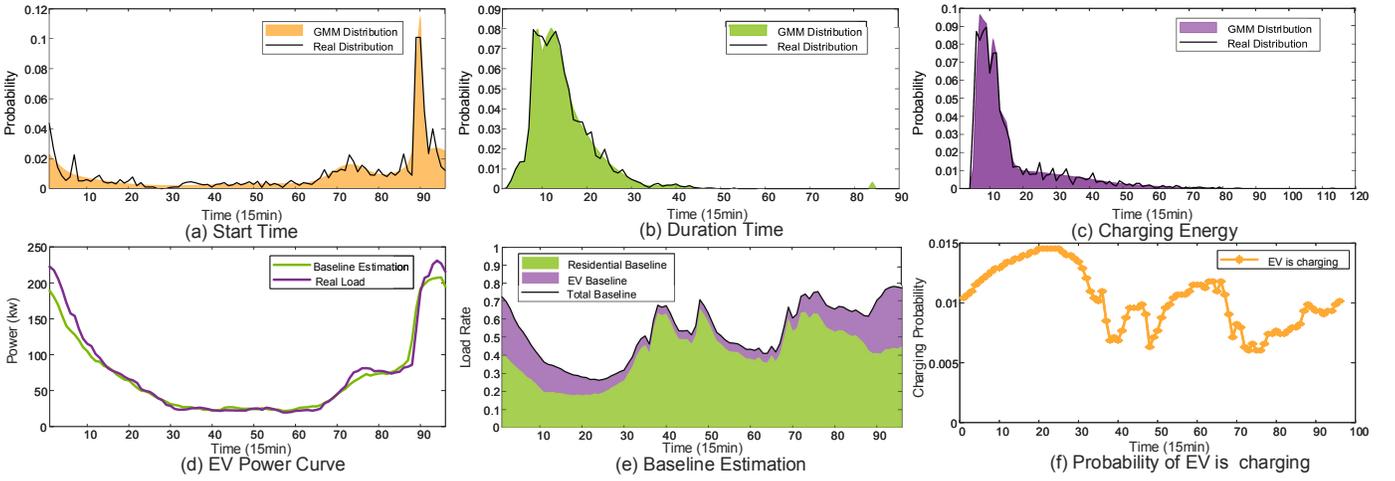} 
	\caption{GMM distribution fitting, baseline load and charging margin estimation} 
	\label{fig2} 
	\vspace{-1.0em}
\end{figure*}
\begin{table}[t]
	\renewcommand{\arraystretch}{1.2}
	\setlength\tabcolsep{8pt}
	\caption{Performance of Different Control Methods under Homogeneous Demand}
	\vspace{-1.5em}
	\begin{center}
		\begin{tabular}{c c c c c}
			\hline
			\hline
			Type &Peak(kW) &$\Delta$Peak$\%$ &PVD(kW)&$\Delta$PVD$\%$\\
			\hline
			Disordered  &600.75 &$\times$ &466.40 &$\times$ \\
			\multirow{2}{*}{\makecell[c]{Centralized \\(Ordered)}}  
			&\multirow{2}{*}{\makecell[c]{413.07 }}
			&\multirow{2}{*}{\makecell[c]{-31.2\% }}
			&\multirow{2}{*}{\makecell[c]{216.02}}
			&\multirow{2}{*}{\makecell[c]{-53.7\%  }}\\
			&&&&\\
			Distributed &463.03 &-22.9\%  &295.01 &-36.7\% \\
			\hline
			\hline
		\end{tabular}
		\label{table6}
	\end{center}
	\vspace{-1.5em}
\end{table}
\begin{figure}[t] 
	\centering 
	\includegraphics[width=0.9\linewidth]{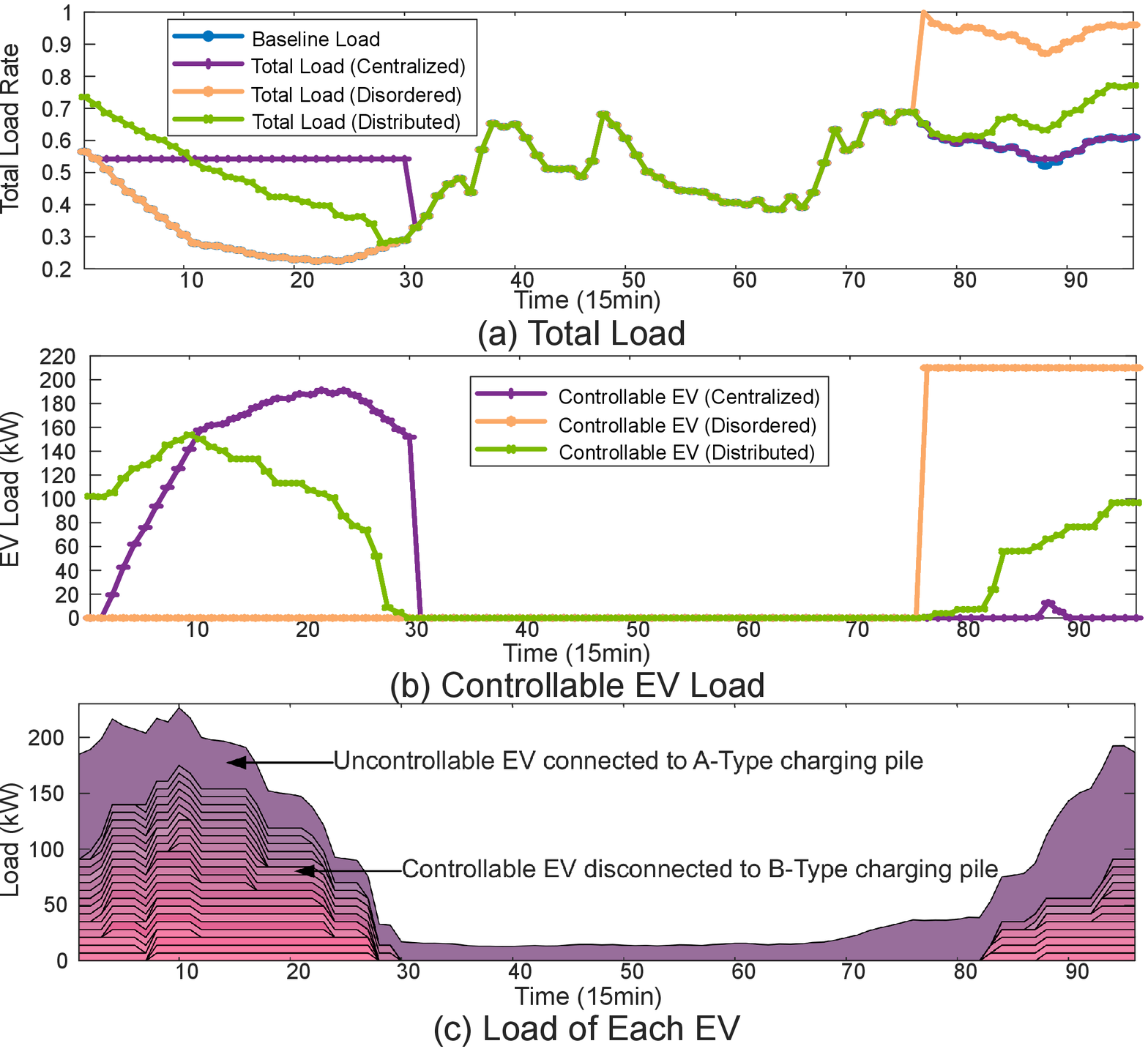} 
	\caption{Comparison of different control methods under homogeneous demand} 
	\label{fig7} 
	\vspace{-1.0em}
\end{figure}
\begin{table}[t]
	\renewcommand{\arraystretch}{1.2}
	\setlength\tabcolsep{8pt}
	\caption{Performance of Different Control Methods under Non-Homogeneous Demand}
	\vspace{-1.5em}
	\begin{center}
		\begin{tabular}{c c c c c}
			\hline
			\hline
			Type &Peak(kW) &$\Delta$Peak$\%$ &PVD(kW)&$\Delta$PVD$\%$\\
			\hline
			Disordered  &476.07 &$\times$ &341.73 &$\times$ \\
			\multirow{2}{*}{\makecell[c]{Centralized \\(Ordered)}}  
			&\multirow{2}{*}{\makecell[c]{413.07 }}
			&\multirow{2}{*}{\makecell[c]{-13.3\% }}
			&\multirow{2}{*}{\makecell[c]{194.35}}
			&\multirow{2}{*}{\makecell[c]{-43.1\%  }}\\
			&&&&\\
			Distributed &417.80 &-12.2\%  &255.25 &-25.3\% \\
			\hline
			\hline
		\end{tabular}
		\label{table2}
	\end{center}
	\vspace{-1.5em}
\end{table}
\begin{figure}[t] 
	\centering 
	\includegraphics[width=0.9\linewidth]{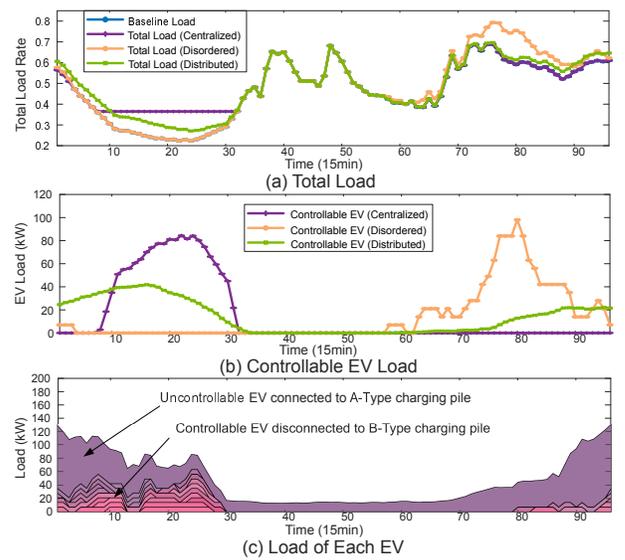} 
	\caption{Comparison of different control methods under non-homogeneous demand} 
	\label{fig4} 
	\vspace{-1.0em}
\end{figure}

Fig. \ref{fig2}(a) - \ref{fig2}(c) illustrate the effect of GMM in fitting the distribution of charging behaviors with start time $\alpha_i$, duration time $\beta_i$ and target charging energy $e_i$. Fig. \ref{fig2}(d) compares the results with the original load. It can be observed that the error between the two curves is sufficiently small, and this method achieves a good estimation. It is worth mentioning that the EV load increases slightly at 20:00, while it increases rapidly at 22:00. At 20:00, few residents return home to start charging, but at 22:00, the electricity price is switched from peak to valley, which leads to a large number of residents choosing to charge EVs at this time to enjoy a lower price which is shown in Table \ref{table7}, but it causes a new load peak. 
\subsection{Comparison of Different Control Methods}
In this section, we observe the comparison results between the method proposed in this paper and the conventional ordered charging control method under different types of EV charging demands and different types of residential load levels. In the scenario of this section, the communication infrastructure failure type is that all B-Type charging piles are disconnected, and the proportion of the number of A-Type charging piles and B-Type charging piles are set to 1:1. It is worth noting that the disordered charging method means that all EVs are uncontrolled and start charging at the time chosen by the drivers according to their wishes, while centralized (or ordered) charging strategy means that when the communication infrastructure is unblocked, all controllable EVs accept the charging commands computed by the central controller.

\subsubsection{Result for Different EV Charging Demand} 
The demand parameters of EV are provided in Table \ref{table4}. It is worth explaining that taking $T^{\rm{arr}}$$\sim$$N(77, 8)$ as an example, it means that $T^{\rm{arr}}$  follows the normal distribution with the mean value of 77 (Equal to 19:15 pm) and the standard deviation of 8 (Equal to 2 hours). Meanwhile, the residential load level is medium and the baseline estimation of residential areas is shown in Fig. \ref{fig2}(e). Fig. \ref{fig2}(f) shows the charging probability of each controllable EV. We compare the proposed algorithm with the disordered charging method and the centralized charging strategy.

\emph{Homogeneous Demand}: Fig. \ref{fig7}(a) and \ref{fig7}(b) illustrate the total load and EV load under three different methods. It can be seen that both the centralized method and the method proposed in this paper can reduce the peak load and PVD. Table \ref{table6} shows the specific numerical results. Although the effect of the method proposed in this paper is not as good as that of the centralized method, it can still play a role in reducing and transferring the peak load when the users disconnected from the utility. Meanwhile, the effect is better than that of the disordered charging. It is clearly seen from Fig. \ref{fig7}(b) that after all EVs arrive at the same time, they charge in disorder together, which causes great pressure on the transformer. In addition, Fig. \ref{fig7}(c) shows the charging behavior of each EV using the proposed method and uncontrollable EV baseline. As a result, Fig. \ref{fig7} shows the advantages of the proposed method, that is, after receiving the transformer capacity margin broadcast by the utility once a day, the EV user does not need to maintain the two-way communication with the utility as the centralized method does. It only needs to independently distribute the calculation and execute the charging command after obtaining the user's demand. When this algorithm is used by large-scale users, the peak load of the transformer can be significantly reduced.
\begin{figure}[t] 
	\centering 
	\includegraphics[width=0.9\linewidth]{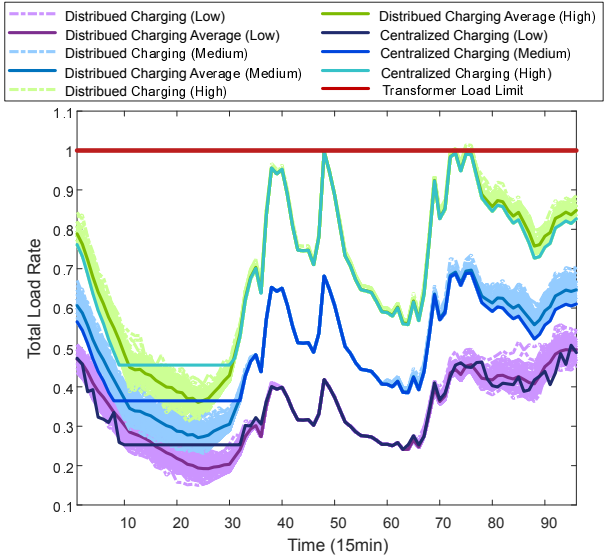} 
	\caption{Comparison of different control methods under different residential load level} 
	\label{fig8} 
	\vspace{-1.0em}
\end{figure}

\emph{Non-Homogeneous Demand}: Fig. \ref{fig4}(a) and \ref{fig4}(b) illustrate the total load and EV load under three different methods and Table \ref{table2} shows the specific numerical results under non-homogeneous demand. It can be seen that the algorithm is still effective when the charging demand of EVs is close to the real situation. In addition, Fig. \ref{fig4}(c) shows the charging behavior of each EV using the proposed method and uncontrollable EV baseline. Besides, the maximum charging probability that the EV is charging is around 5:00 am from Fig. \ref{fig2}(f). But the period of maximum charging power of controllable EVs is not 5:00 am. This is because the proposed method is based on probability sampling, which cannot completely correspond to the charging probability distribution.

\subsubsection{Result for Different Residential Load Level}
In this section, the EV demand is non-homogeneous. We compare the proposed algorithm with the centralized charging strategy under different residential load levels (high, medium, and low). Here, the average residential load rates of high, medium and low levels are 0.70, 0.47, and 0.28.

Fig. \ref{fig8} offers the load curves under different residential load levels using two different algorithms. We can draw the conclusion through observation that under the medium and low levels of residential load, the centralized method and the communication-free distributed method will not exceed the transformer limit. However, when the residential load pressure is high, the centralized method can still provide charging services for all controllable EVs without exceeding the transformer limit. However, the performance of our proposed method is limited and will not ensure that it does not exceed the transformer limit where after 1000 random tests, the overload ratio of the transformer is 13.5\%. It is proved again that the method proposed in this paper is not as effective as the centralized control method, but it can still reduce and transfer the peak load when the users disconnected from the utility.
\begin{table}[t]
	\renewcommand{\arraystretch}{1.2}
	\setlength\tabcolsep{4pt}
	\caption{Parameter Sensitivity Analysis}
	\vspace{-1.5em}
	\begin{center}
		\begin{tabular}{c c c c c c}
			\hline
			\hline
			Controllable EV\% &Peak(kW) &$\Delta$Peak$\%$ &PVD(kW)&$\Delta$PVD$\%$& MAC\\
			\hline
			0\%          &476.07 &$\times$ &341.73 &$\times$  & 87\\
			20\%          &437.36 &-8.1\%   &274.39 &-19.7\%    &117 \\
			40\%          &424.30 &-10.9\%   &254.88 &-24.4\%   &140 \\
			60\%          &414.22 &-13.0\%  &249.80 &-26.9\%   &163 \\
			80\%          &402.95 &-15.4\%  &237.91 &-30.4\%   & 180\\
			100\%        &396.79 &-16.7\%  &229.32 &-32.9\%   &214 \\
			\hline
			\hline
		\end{tabular}
		\label{table3}
	\end{center}
	\vspace{-1.5em}
\end{table}
\subsection{Sensitivity Analysis of Different Proportion Charging Pile Types}
In this section, we mainly compare the impact of different proportions of controllable EVs in the residential area on the peak load, PVD, and the maximum acceptable capacity (MAC) of EVs under the condition of communication infrastructure failure. It can be seen from Table. \ref{table3} that: i) The peak and PVD of the total load are reduced due to the implementation of the distributed charging strategy. ii) With the increase in the proportion of controllable EVs, the peak and PVD of load decrease even more. iii) When EVs are uncontrollable, the maximum number of EVs that can be accommodated in the residential area is exactly the number of users. iv) When the proportion of controllable EVs increases, the MAC of EVs in the residential area increases. When all residential areas are controllable EVs, the MAC of EVs in the residential area is 2.4 times that when they are uncontrollable.
\subsection{Influence of Different Proportions of Communication Failures}
In this section, we observed the results of different proportions of B-Type charging piles partially disconnected. In the scenario of this section, the EV charging demand type is non-homogeneous demand, and the residential load level is medium level. All the charging piles are the B-Type charging piles.

\begin{figure}[t]
	\centering 
	\includegraphics[width=0.9\linewidth]{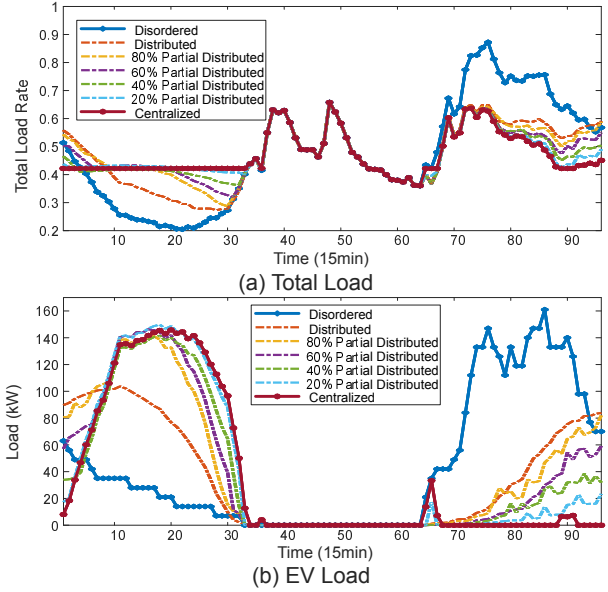} 
	\caption{Comparison of different control methods with partial disconnection} 
	\label{fig3.6} 
	\vspace{-1.0em}
\end{figure}

The test results of this section are shown in Fig. \ref{fig3.6} and Table \ref{table3.6}. It can be observed from Fig. \ref{fig3.6} that the method proposed in this paper can significantly reduce the peak load and PVD of the transformer when all B-Type charging piles are disconnected or the B-Type charging piles are partially disconnected. Meanwhile, in this case, we can see that the proposed method in this article has similar performance in reducing peak load compared to the centralized charging method. However, from Fig. \ref{fig3.6}, it can be observed that the PVD when all the B-Type charging piles are disconnected is significantly greater than when they are partially disconnected, which proves that the performance of the method proposed in this paper is still slightly inferior to that of the centralized method. Therefore, the proposed method is a scheme to reduce the cost and risk of communication equipment, which conforms to our original intention.
\subsection{Computation Efficiency Using Edge Modules with Different Resources}
In this section, we compared the computational efficiency of edge modules with different resources. The computer is used to validate algorithms, Raspberry Pi is used to emulate EVs, and the MCU is used to emulate smart meter. Table \ref{table4.7} first displays the basic information such as the device type, function, and solution method that we are using. We also conducted two sets of tests on different devices. From Table \ref{table4.7}, it can be observed that in both computer and Raspberry Pi, regardless of whether a solver is used or not, the algorithm can quickly make plans for EV charging for up to 96 time periods per day. In the MCU, with scarce computing resources (Due to cost constraints and insufficient computing resources, it is difficult to use solvers), this paper can still quickly carry out charging planning for up to 12 time periods per day, which can meet practical needs and has application value.
\vspace{-1.0em}
\section{Conclusion}\label{6}
This paper has proposed a communication-free distributed charging control scheme as a backup to provide a fully decentralized on-site charging strategy for EV group after a communication failure. Our scheme is able to estimate the charging margin for controllable EVs considering charging behavior modeled by GMM. In this scheme, the distributed control algorithm that we design for controllable EVs is based on probabilistic capacity margin, so that the user can autonomously formulate the charging plan of the EV without the help of communication infrastructure. Based on the real residential load of a residential area in Hangzhou, our algorithm can reduce the peak of the transformer by up to about 16\%-25\% and the PVD by up to about 30\%-45\%. Its performance is superior to disordered charging, but weaker than the centralized algorithm when the communication link is reliable. As a result,  in case of communication failure, the probability of the EV group charging out of control can be properly reduced. In the future, we will consider using machine learning- or reinforcement learning-based method and extend the algorithm to how to combine the user's past charging habits and electricity price with the capacity margin so that the user can still autonomously formulate a charging strategy in case of communication failure and loss of connection between the user and the utility.
\begin{table}[t]
	\renewcommand{\arraystretch}{1.2}
	\setlength\tabcolsep{4pt}
	\caption{Performance of Different Control Methods with Partial Disconnection}
	\vspace{-1.5em}
	\begin{center}
		\begin{tabular}{c c c c c}
			\hline
			\hline
			\multirow{2}{*}{\makecell[c]{Proportion of \\Disconnection\%  }}&\multirow{2}{*}{\makecell[c]{Peak(kW)  }}&\multirow{2}{*}{\makecell[c]{$\Delta$Peak$\%$  }}&\multirow{2}{*}{\makecell[c]{PVD(kW) }}
			&\multirow{2}{*}{\makecell[c]{$\Delta$PVD$\%$ }}\\
			&&&&\\
			\hline
			Disordered &523.34&$\times$&400.42&$\times$\\
			0\%            &394.55 &-24.6\%&177.07 &-55.8\% \\
			20\%          &394.55&-24.6\% &177.07&-55.8\%  \\
			40\%          &394.55&-24.6\%  &177.07 &-55.8\% \\
			60\%          &394.55&-24.6\%&202.55 &-49.9\%  \\
			80\%          &394.55&-24.6\%&222.49 &-44.3\%   \\
			100\%        &394.55&-24.6\%&230.61 &-42.4\%   \\
			\hline
			\hline
		\end{tabular}
		\label{table3.6}
	\end{center}
	\vspace{-1.5em}
\end{table}
\begin{table*}[t]
	\renewcommand{\arraystretch}{1.2}
	\setlength\tabcolsep{8pt}
	\caption{Computational Efficiency of Algorithm on Devices with Different Resources}
	\vspace{-1.5em}
	\begin{center}
		\begin{tabular}{c c c c c}
			\hline
			\hline
			\multirow{2}{*}{\makecell[c]{Device Type (Core)}} &\multirow{2}{*}{\makecell[c]{Purpose}} &\multirow{2}{*}{\makecell[c]{Time Period}} &\multicolumn{2}{c}{Run Time (s)}\\
			\cline{4-5}
			&&&\makecell[c]{PC-IPM (External Solver Ipopt)}
			&\makecell[c]{PD-IPM  (Handed Code)}\\
			\hline
			Computer (Intel Xeon X5650)&Algorithm Validation&96&1.3936&3.7660\\ 
			Raspberry Pi 3B (ARM Cortex A53)&EV Emulation &96&1.1591 &7.1041\\
			MCU-STM32F429 (ARM Cortex M4)&Smart Meter Emulation &12&$\sim$&12.8838\\
			\hline
			\hline
		\end{tabular}
		\label{table4.7}
	\end{center}
	\vspace{-1.5em}
\end{table*}
\section*{Appendix A}\label{AppA}
The determinant form of $\boldsymbol{\Gamma}_1$ in (\ref{eq2022110204}) can be expresses as (\ref{eq2023061201}):
\begin{equation}
	\setlength{\abovedisplayskip}{3pt} 
	\setlength{\belowdisplayskip}{3pt}
	\setlength{\arraycolsep}{5pt} 
	\boldsymbol{\Gamma_1} =
\overbrace{
	\begin{bmatrix}
		\begin{matrix}
			1 &0 &... &0 &0 &0 \\
			1 &1 &0 &... &0 &0  \\
			1 &1 &1 &0 &... &0 \\
			1 &1 &1 &1 &0 &... \\
			0 &1 &1 &1 &1 &0 \\
			... &... &... &... &... &... 
		\end{matrix} &
		\begin{matrix}
			0 &1 &1 &1\\
			0 &0 &1 &1 \\
			0 &0 &0 &1\\
			0 &0 &0 &0\\
			... &0 &0 &0\\
			... &... &... &...
		\end{matrix} \\
		\begin{matrix}
			0 &... &0 &1 &1 &1\\
			0 &0 &... &0 &1 &1 \\
			0 &0 &0 &... &0 &1 \\
			0 &0 &0 &0 &... &0 
		\end{matrix} & 	
		\underbrace{	
			\begin{matrix}
				1 &0 &0 &0\\
				1 &1 &0 &0 \\
				1 &1 &1 &0\\
				1 &1 &1 &1
		\end{matrix}}_{\beta^{'}}
	\end{bmatrix}
}^{H}
\label{eq2023061201}
\end{equation}
\section*{Appendix B}
The standard form of the QP problem with constraints is shown in (\ref{eq2023030401}) and (\ref{eq2023030402}):
\begin{equation}
	\setlength{\abovedisplayskip}{3pt} 
	\setlength{\belowdisplayskip}{3pt}
	\min\enskip f(\boldsymbol{x}) = {1\over 2}\boldsymbol{x}^T\boldsymbol{B} \boldsymbol{x} +  \boldsymbol{c}^T\boldsymbol{x}
	\label{eq2023030401}
\end{equation}
\begin{equation}
	\setlength{\abovedisplayskip}{3pt} 
	\setlength{\belowdisplayskip}{3pt}
	\rm{s.t.} \enskip\boldsymbol{A}\boldsymbol{x}\geq\boldsymbol{b}\label{eq2023030402}
\end{equation}
where $\boldsymbol{B}$ is an $n \times n$ symmetric matrix. $\boldsymbol{A}$ is an $m \times n$ matrix. $\boldsymbol{c}$ and $\boldsymbol{b}$ are respectively $n$-dimensional and $m$-dimensional column vectors. 

After converting the original problem which is shown in (\ref{eq2023021011}) and (\ref{eq2023021010}) into the standard form shown in (\ref{eq2023030401}) and (\ref{eq2023030402}), the form of the matrix $\boldsymbol{B}$ can be expressed in (\ref{eq2023030403}):

\begin{equation}
	\setlength{\abovedisplayskip}{3pt} 
	\setlength{\belowdisplayskip}{3pt}
	\setlength{\arraycolsep}{0.8pt} 
	 \boldsymbol{B} =
	 	\overbrace{
	 	\begin{bmatrix}
	 	\beta^{'}     &\beta^{'}-1 &\beta^{'}-2&...                 &0 &...&\beta^{'}-2&\beta^{'}-1 \\
	 	\beta^{'}-1 &\beta^{'}     &\beta^{'}-1&...                 &0 &...&\beta^{'}-3&\beta^{'}-2 \\
	 	\beta^{'}-2 &\beta^{'}-1 &\beta^{'}    &\beta^{'}-1&...&0&...&\beta^{'}-3 \\
	 	...                  &...                   &...                   &...  &...  &... &...  &...\\
	 	\beta^{'}-3 &...&0  &...  &\beta^{'}-1 &\beta^{'} &\beta^{'}-1  &\beta^{'}-2 \\
	 	\beta^{'}-2 &\beta^{'}-3&...  &0  &...  &\beta^{'}-1 &\beta^{'}  &\beta^{'}-1\\
	 	\beta^{'}-1 &\beta^{'}-2&...  &0  &...  &\beta^{'}-2 &\beta^{'}-1  &\beta^{'} 
	 \end{bmatrix}
	 }^{H}
	\label{eq2023030403}
\end{equation}

According to (\ref{eq2023030401}), ${1\over 2}\boldsymbol{x}^T\boldsymbol{B} \boldsymbol{x}$ is a real symmetric matrix. Meanwhile, all elements in the matrix (\ref{eq2023030403}) are not less than $0$. In addition, according to the constraint (\ref{eq2023030401}) and equation (\ref{eq2023021010}), all elements in vector $\boldsymbol{x}$ are greater than $0$. Therefore, the term ${1\over 2}\boldsymbol{x}^T\boldsymbol{B} \boldsymbol{x}$ is never less than $0$. According to the definition of positive semidefinite quadratic form and positive semidefinite matrix, the term  ${1\over 2}\boldsymbol{x}^T\boldsymbol{B} \boldsymbol{x}$ is positive semidefinite quadratic form and matrix $\boldsymbol{B}$ is positive semidefinite matrix, which is equivalent to the fact the QP problem shown in (\ref{eq2023021011}) and (\ref{eq2023021010}) is solvable.
\bibliographystyle{IEEEtran}
\bibliography{IEEEabrv,IEEEexample}
\begin{IEEEbiographynophoto}{Heyang Yu} (S'19) received his B.S. degree in electrical engineering from the College of Electrical Engineering, Zhejiang University, Hangzhou, China, in 2019. 
	
Currently, he is pursuing the Ph.D. degree at the College of Electrical Engineering, Zhejiang University, Hangzhou, China. His research interests include non-intrusive load monitoring and residential demand response.
\end{IEEEbiographynophoto}
\begin{IEEEbiographynophoto}{Chuanzi Xu} received her B.S. and M.S. degrees in electrical engineering from the College of Electrical Engineering, Zhejiang University, Hangzhou, China, in 2008 and 2017, respectively. 
	
Currently, she is the executive deputy director of Integration Innovation Center of Hangzhou Power Supply Company of State Grid Zhejiang Electric Power Co. Ltd., Hangzhou, China. Her research interests include big data analytics in power systems and emerging business scheme development on customer sides.
\end{IEEEbiographynophoto}
\begin{IEEEbiographynophoto}{Weifeng Wang} received his B.S. degree in electrical engineering from the College of Information Engineering, Zhejiang University of Technology and the M.S. degree in electrical engineering from the College of Electrical Engineering, Zhejiang University, Hangzhou, China, in 2000 and 2013, respectively. 
	
He is currently deputy director of Measurement Division of Marketing Department of State Grid Zhejiang Electric Power Co. Ltd., Hangzhou, China. His research interest is advanced metering infrastructure.
\end{IEEEbiographynophoto}
\begin{IEEEbiographynophoto}{Guangchao Geng} (S'10-M'14-SM'19) received his B.S. and Ph.D. degrees in electrical engineering from the College of Electrical Engineering, Zhejiang University, Hangzhou, China, in 2009 and 2014, respectively. From 2012 to 2013, he was a visiting student at the Department of Electrical and Computer Engineering, Iowa State University, Ames, United States. From 2014 to 2017, he was a post-doctoral fellow at the College of Control Science and Engineering, Zhejiang University, Hangzhou, China and the Department of Electrical and Computer Engineering, University of Alberta, Edmonton, AB, Canada. He joined the faculty of College of Electrical Engineering, Zhejiang University in 2017. 
	
Currently, he is an associate professor at the College of Electrical Engineering, Zhejiang University, Hangzhou, China. His research interests include non-intrusive sensing technology, data analytics in power systems, power system stability and control.
\end{IEEEbiographynophoto}
\begin{IEEEbiographynophoto}{Quanyuan Jiang}(M'10-SM'19) received his B.S., M.S., and Ph.D. degrees in electrical engineering from Huazhong University of Science \& Technology, Wuhan, China in 1997, 2000, and 2003, respectively. From 2006 to 2008, he was a visiting associate professor at the School of Electrical and Computer Engineering, Cornell University, Ithaca, United States.
		
He is currently a professor at the College of Electrical Engineering and academic dean of Graduate School, Zhejiang University, Hangzhou, China. His research interests include power system stability and control, applications of energy storage systems and high performance computing in power systems.
\end{IEEEbiographynophoto}
\end{document}